\documentclass[12pt]{iopart}
\usepackage{iopams}  
\usepackage{citesort}
\usepackage{doi}
\usepackage{cite}
\usepackage{graphicx}
\usepackage{subfig}
\newcommand{\eq}{\begin{equation}}
\newcommand{\qe}{\end{equation}}

\newcommand{\mbf}[1]{\mathrm{#1}}

\newcommand{\ket}[1]{\left| #1\right>}
\newcommand{\inpro}[2]{\left< #1 \right|\left. #2 \right>}
\newcommand{\exd}{{\rm d}}
\newcommand{\be}{\begin{equation}}
\newcommand{\ee}{\end{equation}}
\newcommand{\commutator}[2]
{
 [ #1 ,  #2
  ]
}
\newcommand{\pref}[1]{(\ref{#1})}

\begin{document}

\title[]{Duality between the quantum inverted harmonic oscillator and inverse square potentials}

\author{Sriram Sundaram\textsuperscript{1}, C. P. Burgess\textsuperscript{1,2,3}, D. H. J. O'Dell\textsuperscript{1}}

\address{$^{1}$Department of Physics and Astronomy, McMaster University, 1280 Main Street W., Hamilton, Ontario, Canada, L8S 4M1}
\address{$^{2}$Perimeter Institute for Theoretical Physics, 
Waterloo, Ontario, Canada, N2L 2Y5}
\address{$^{3}$ School of Theoretical Physics, Dublin Institute for
 Advanced Studies, 10 Burlington Rd., Dublin,
 Co. Dublin, Ireland}
\ead{dodell@mcmaster.ca}
\vspace{10pt}
\begin{indented}
\item[]
\end{indented}
\begin{abstract}

In this paper we show how the quantum mechanics of the inverted harmonic oscillator can be mapped to the quantum mechanics of a particle in a super-critical inverse square potential. We demonstrate this by relating both of these systems to the Berry-Keating system with hamiltonian $H=(xp+px)/2$. It has long been appreciated that the quantum mechanics of the inverse square potential has an ambiguity in choosing a boundary condition near the origin and we show how this ambiguity is mapped to the inverted harmonic oscillator system. Imposing a boundary condition requires specifying a distance scale where it is applied and changes to this scale come with a renormalization group (RG) evolution of the boundary condition that ensures observables do not directly depend on the scale (which is arbitrary). Physical scales instead emerge as RG invariants of this evolution. The RG flow for the inverse square potential is known to follow limit cycles describing the discrete breaking of classical scale invariance in a simple example of a quantum anomaly, and we find that limit cycles also occur for the inverted harmonic oscillator. However, unlike the inverse square potential where the continuous scaling symmetry is explicit, in the case of the inverted harmonic oscillator it is hidden and occurs because the hamiltonian is part of a larger su(1,1) spectrum generating  algebra.  Our map does not require the boundary condition to be self-adjoint, as can be appropriate for systems that involve the absorption or emission of particles. 
  
\end{abstract}

%
\vspace{2pc}
\noindent{\it Keywords}: inverted harmonic oscillator, inverse square potential, duality, renormalization group
%
%
%
%

\section{Introduction}

The inverted harmonic oscillator (IHO) describes a particle moving in a potential $V_{\mathrm{IHO}}(x) \propto -x^2$ which is singular at infinity, whereas the attractive inverse-square potential (ISP) varies as $V_{\mathrm{ISP}}(x) \propto -1/x^2$ and is singular at the origin. These two well-studied systems are generic models for unstable and scale-invariant systems, respectively, and at first sight seem to give rise to opposite behaviour because the IHO potential drives particles to large $x$, whereas the ISP causes ``fall to the center''  \cite{lifshitz1965quantum, perelomov1970faltothecenter}. In this paper we demonstrate how in fact the Hamiltonians and quantum states of these systems can be explicitly mapped into one another, showing them to be in some sense alternative descriptions of equivalent physics. In so doing we also relate both of these models to the Berry-Keating (BK) system which has the classical Hamiltonian $xp$ which in quantum mechanics we symmeterize to  $H=(xp+px)/2$ to make it formally hermitian (but not necessarily self-adjoint). Although the canonical transformation that relates the IHO and BK models has been extensively studied before \cite{balazs1990wigner, brout1995primer, berry1999h, chruscinski2003quantum, subramanyan2021physics, ullinger2022logarithmic}, we believe that the connection to the ISP is new.  

The quantum mechanics of the IHO is exactly solvable \cite{barton1986quantum} and appears in many branches of physics where it provides a simple prototype for  instability or tunneling through a smooth barrier \cite{yuce2006inverted,subramanyan2021physics,bhattacharyya2021multi}. Particular examples include the Landau-Zener model in atomic and molecular physics \cite{landau1932,zener1932}, squeezed states, amplifiers, and the Glauber oscillator  in quantum optics \cite{yurke1986, prakash1994mixed,glauber1986amplifiers,gentilini2015physical}, the quantum hall effect in condensed matter physics \cite{subramanyan2021physics}, non-equilibrium phase transitions in statistical physics \cite{gietka2021inverted}, studies of chaos and complexity \cite{qu2022chaos, hashimoto2020exponential}, and  Riemann zeroes \cite{bhaduri1997riemann}. 
In quantum field theory the IHO arises in Schwinger pair production  \cite{brout1995primer,srinivasan1999particle}, Hawking radiation from black holes \cite{schutzhold2013hawking}, squeezing of states in inflationary cosmology \cite{albrecht1994inflation}, tachyon physics \cite{das2008origin}, is widely used in string theory \cite{dhar1995discrete, klebanov2003d, maldacena2005flux, sen2023infrared},  and so on.

The ISP likewise arises in multiple scenarios. It occurs as the interaction potential between an electron and a neutral polar molecule \cite{levyleblond67,camblong2001} (or similarly between a charged wire and an atom \cite{schmiedmayer1998coldatom,plestid2018fall}), as an effective description for three-body bound states in the Efimov effect \cite{EFIMOV1970,EFIMOV1973,werner06,BRAATEN2006,bhaduri2011efimov,moroz2015efimov}  that was originally predicted in nuclear physics  and has been studied experimentally in detail using ultracold atoms \cite{kramer2006,williams2009,pollack2009,huang2014,tung2014}, in statistical mechanics through the exactly solvable Calogero-Sutherland quantum many-body problem with pairwise ISPs \cite{calogero1969,sutherland1971,pkp1999equivalence,gurappa2000free} and as a model for winding transitions relevant to polymers such as DNA \cite{nisoli2014}, in the study of coherence in optics \cite{sundaram2016origin}, in supersymmetric quantum mechanics \cite{gangopadhyaya1994analysis}, in the AdS/CFT correspondence \cite{moroz2010AdsCFT}, and in the near-horizon physics of black holes \cite{srinivasan1999particle,Gupta2001horizon,camblong2003,burgess2018effective}. 

A key feature of the ISP is that the non-relativistic Schr\"{o}dinger equation with this potential is scale invariant as both the kinetic energy and  potential terms scale as length$^{-2}$ and thus there is no natural length scale present (such as the Bohr radius in the Coulomb problem, say). However, if the singular nature of the ISP at the origin is tamed by introducing a cut-off or boundary condition at short distance (in the physical examples given above the ISP only models the long wavelength behaviour), this regulator necessarily breaks the scale invariance in a simple example of a quantum anomaly \cite{callan1970,colemanbook,holstein2002anomalies,essin2006,olshanii2010anomaly}. Furthermore, in order to ensure that the long wavelength physics is independent of the regulator the theory's couplings should be renormalized \cite{gupta1993renormalization}. In the case of the super-critical ISP (where the strength of the potential overcomes the zero-point energy) this leads to a characteristic renormalization group (RG) flow that takes the form of limit cycles   \cite{bawin2003invsqrg,mueller2004renormalization,kaplan09,moroz2010nonrelativistic}. In a previous paper on these limit cycles, we emphasized that the change from sub- to super-criticality is a type of $\mathcal{PT}$ symmetry breaking transition where the fixed points of the RG flow change from real valued, describing unitary physics, to a complex conjugate pair, one describing pure emission and the other pure absorption \cite{sriram_symmetry} (see also \cite{stalhammar2023}).

 The third system in our trio is the BK Hamiltonian. It has been extensively studied in the context of quantum chaos and attempts to prove the Riemann hypothesis \cite{berry86,berry1999h,connes99,melintransform_xp,srednicki2011,bender2017,sierra2019}. 
The dynamics generated by the classical BK Hamiltonian is exponentially unstable, unbounded, and breaks time reversal symmetry \cite{berry1999h}.
Although it does not have distinct kinetic and potential energies, the BK Hamiltonian is manifestly scale invariant, like the ISP.  
A particular way of modifying the BK Hamiltonian exhibits a cyclic RG flow similar to the ISP and has been used to map to ``Russian-doll'' models of superconductivity \cite{sierra2005riemann}. It has also been argued to capture aspects of  black hole physics \cite{dalui2020near, dalui2019presence}. A Dirac-type variant of the BK model 
in two dimensions in which the operator $p$ is replaced by $\sigma . p$ has been proposed by Gupta \etal \cite{gupta2013dirac} and shown to be equivalent to a Schr\"{o}dinger equation with an ISP and an additional Coulomb potential. This model finds physical applications in describing gapped graphene with a super-critical Coulomb charge. In fact, the Dirac equation for a massless fermion in the presence of an attractive Coulomb potential is also scale invariant because both terms scale as $1/r$ and the quantum anomaly that breaks this continuous scaling symmetry has been observed in an experiment on graphene \cite{anomalygraphene2017ovdat}. 

In the duality scheme we lay out in this paper, the BK model provides a stepping stone between the IHO and the ISP.  In the first step the BK Hamiltonian is obtained from the IHO via a canonical transformation, and in the second step the Schr\"{o}dinger equation with an ISP is reached by squaring the BK Hamiltonian and then applying an integrating factor to remove a first order derivative. Since the exact solutions of the Schr\"{o}dinger equations defining the IHO, BK,  and ISP models are all already known (parabolic cylinder functions, monomials, and confluent hypergeometric functions, respectively) the difficulty in treating these models does not lie in finding solutions to differential equations. Rather, it lies in choosing the correct boundary conditions that these solutions must obey especially in view of the fact that their energies form a continuum and are unbounded from below.  One of the novelties of the present paper  therefore lies in mapping the boundary conditions between the models and exhibiting how they behave under renormalization.
 
 \par The rest of this paper is organized as follows : after putting the current work in a larger historical context in  \Sref{sec:dualities}, \Sref{sec:systems} describes the three dual systems, their eigenfunctions and symmetries. The need for appropriate far field physics (boundary condition at long distances) for the quantum mechanics of an IHO is discussed. Using a canonical transformation we map the Schr\"{o}dinger equation with an IHO potential in one set of variables $\xi$, to a Schr\"{o}dinger equation with a super-critical ISP in another set of variables $Q$, which, however, now has ambiguities in fixing a boundary condition for the wavefunction near the origin. This problem can be tackled systematically using point particle effective field theory (PPEFT) which suggests a general linear (Robin) form for the boundary condition. 
In \Sref{sec:mapping} we apply this boundary condition for the wavefunction near the origin of the ISP in an RG invariant way. Furthermore, we do a one-to-one mapping of the inverse square states to the IHO states using a quantum canonical transform. For the IHO problem, the boundary condition for the asymptotic parabolic cylinder functions is also fixed by a linear boundary condition, but now at large distances, and also in an RG invariant way. Conclusions are given in \Sref{sec:conclusions}. We also include three appendices that discuss the properties of parabolic cylinder functions, quantum canonical transformations, and other details needed for the mappings.

\section{A brief history of power law dualities}
\label{sec:dualities}

To put the current paper in context it is worth mentioning the history of  dualities between power law potentials (for fascinating reviews of this topic, that goes to the very foundations of modern physics, see \cite{arnold1990huygens} and \cite{needham1993}). The most famous duality is the relation  between classical motion in a gravitational potential $V_{\mathrm{grav}} \propto -1/r$ and a (stable) planar harmonic oscillator potential $V_{\mathrm{HO}} \propto r^2$, which are associated with the names of Newton and Hooke, respectively. Both give rise to closed orbits which are ellipses: in the harmonic oscillator case the force centre is located at the centre of the ellipse whereas in the gravitational case the force centre is at a focus. The two cases can be mapped onto each other by \textit{squaring} the harmonic oscillator ellipse to obtain the gravitational ellipse, as described by K. Bohlin in 1911 \cite{bohlin1911,horvathy2014shotokepler} (we note that the mapping between the IHO and the ISP to be discussed in this paper also involves a step where the Hamiltonian is squared). 
In fact, there is a continuous family of dual potentials $V \propto r^{\alpha} \leftrightarrow V \propto r^{\overline{\alpha}}$ determined by the relation \cite{kasner1913,arnold1989,grant1994classical}
\begin{equation}
(\alpha+2)(\overline{\alpha}+2)=4
\label{eq:kasner}
\end{equation}
of which the harmonic oscillator-gravity duality $(\alpha,\overline{\alpha})=(-1,2)$ is only one example. The other integer-valued cases are $(-3,-6)$, $(-4,-4)$, and $(0,0)$ (the last case can be interpreted as corresponding to a logarithmic potential). This duality relation was derived by E. Kasner in 1913 \cite{kasner1913}, but seems to have also been included in the 1720 treatise by C. MacLaurin \cite{albouy2022}, and Newton discussed the self-dual cases $(-4,-4)$ and $(0,0)$ in his Principia \cite{arnold1990huygens,newton1687,newton1999,chandrasekhar2003newton}.  Extensions to quantum mechanics and other generalizations have been also been widely studied, see for example \cite{faure1953,bateman1992}. A case which is particularly relevant in current ultracold atom physics is the equivalence between free quantum particles and those in harmonic potentials \cite{steuernagel2014} since in experiments the atoms are often confined in harmonic traps but calculations of interacting many-particle systems are of course easier for plane wave states.

Does the duality considered in this paper fit into the above scheme? On the one hand, putting $\alpha=-2$ for the ISP into Eq.\ (\ref{eq:kasner}) makes the left hand side vanish so that $\overline{\alpha}$ is undefined, and on the other hand putting $\alpha=2$ for the IHO gives $\overline{\alpha}=-1$ corresponding to the gravitational case as expected, and so does not seem to include the ISP-IHO duality as a possibility. The result given in Eq.\ (\ref{eq:kasner}) takes no account of the fact that the IHO is inverted and when this is done one can map hyperbolic trajectories between the gravitational and IHO potentials \cite{needham1993}. Proceeding in a slightly different way, Wu and Sprung considered the limiting procedure where $\alpha \rightarrow \pm \infty$ in classical mechanics in two dimensions and have shown that it can represent a hard wall well or hard sphere scattering  and that the dual potential is the ISP \cite{wu_sprung1995}. However, the duality we study here is of a different nature since it maps between small and large distances (see \fref{fig4}) such that fall-to-the-centre in the ISP becomes fall-to-infinity in the IHO. Furthermore, the scale invariance which is such an important part of the quantum behaviour of the ISP is not replicated in the classical mechanics as the kinetic energy in this case does not scale as an inverse square length. We leave it as an open problem as to whether Eq.\ (\ref{eq:kasner}) can be re-interpreted in a creative way that includes the ISP-IHO duality to be detailed below. 

\section{The systems}
\label{sec:systems}

This section gives a brief description of each of the three systems that are to be related. 

\subsection{Inverted harmonic oscillator}
\label{fall_to_infinity}
\par The IHO is defined by the Hamiltonian:
\begin{equation}
H(x,p) = \frac{p^{2}}{2m} - \frac{1}{2}m\omega^{2}x^{2} 
\end{equation}
and so the time-independent Schr\"{o}dinger equation for energy eigenvalue \footnote{Perversely, for later convenience we denote the system energy by $-E$ so that $E>0$ describes negative-energy states.} ${\cal E} = -E$ written in the position representation is: 
\begin{equation}
\left[\frac{-\hbar^{2}}{2m}\frac{\partial^{2}}{\partial x^{2}} -\frac{1}{2}m\omega^{2}x^{2}\right] \phi (x) = -E \phi(x) \ . \label{iho}
\end{equation}
\begin{figure}
\includegraphics{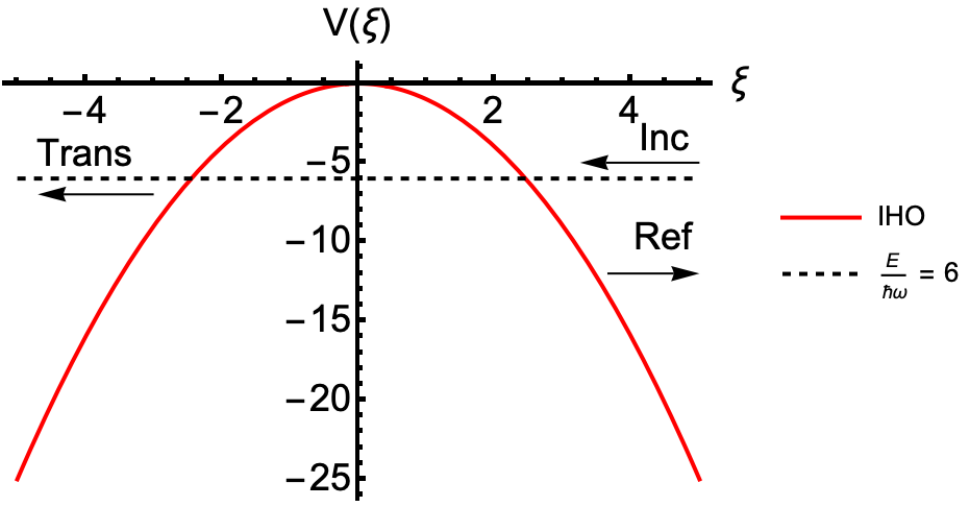}
\caption{The figure shows the IHO potential $V(\xi) = - \frac{1}{2} \xi^{2}$, with a representative negative energy eigenvalue also drawn (corresponding to the choice $\hat{E}= E/\hbar \omega = 6$).}
\label{fig1}
\end{figure} 
It is convenient to use the following dimensionless coordinate
\begin{equation}
\xi = \sqrt{\frac{m\omega}{\hbar}}x   \label{xi_variable}
\end{equation}
in terms of which (\ref{iho}) becomes
\begin{equation}
 \frac{\pi^{2}-\xi^{2}}{2} \phi(\xi) = -\frac{E}{\hbar\omega} \phi (\xi) = - \hat E \label{ihoxipi}
\end{equation}
where $\hat E := E/\hbar\omega$ and the canonical momentum $\pi = -\mbf{i} {\partial}/{\partial \xi}$ satisfies $\commutator{\xi}{\pi} = \mbf{i}$. 
\par The IHO potential is shown in \fref{fig1} together with a representative negative energy eigenvalue. Notice that both positive and negative energy eigenvalues are allowed and instability arises because the spectrum is not bounded from below. For negative energy states ({\it i.e.}~$E > 0$ in our convention) evolution between large negative and positive positions is a tunnelling problem, while for positive energy states ($E < 0$) it is instead a classically allowed barrier scattering problem. 
\par
\begin{figure*}
\graphicspath{{./Plots/}}
\subfloat[]{\includegraphics[width = 0.45\columnwidth]{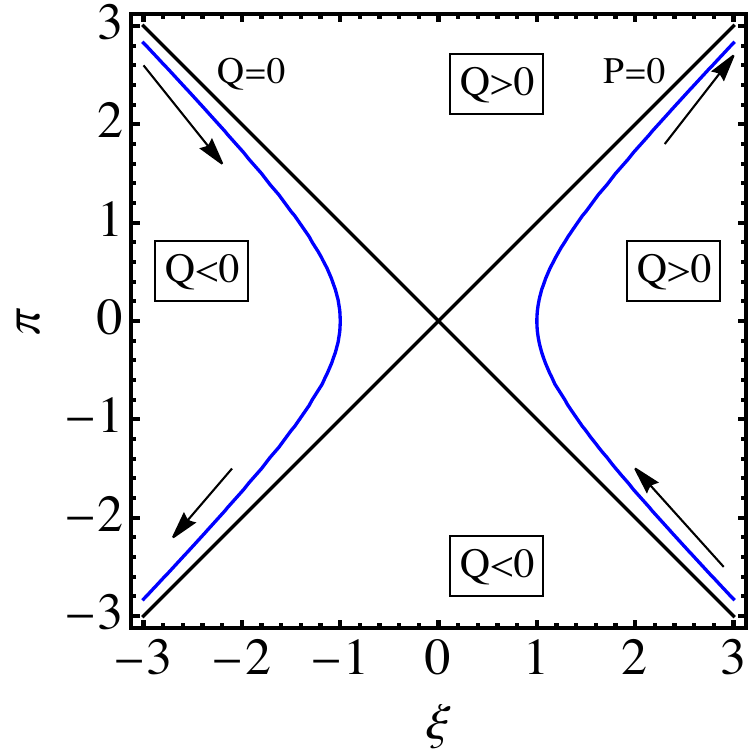}}\hspace{1cm}
\subfloat[]{\includegraphics[width = 0.45\columnwidth]{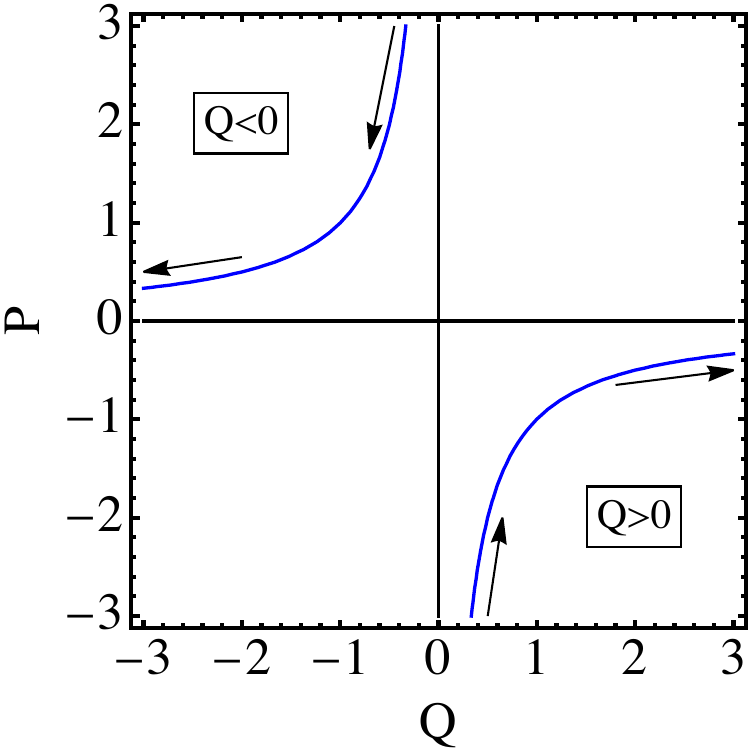}}
\caption{Panel (a) shows the classical phase space portrait of an IHO with negative energy ($E > 0$) with hyperbolic trajectories. Panel (b) shows the same trajectories drawn using the canonically related coordinates $Q$ and $P$ described in the text. The canonical transformation corresponds to a $\pi/4$ rotation in phase space.}
\label{fig2}
\end{figure*}

 The classical turning points for negative energy ($E>0$) are $\xi_{0} = \pm \sqrt{2 \hat{E}}$ and the classical negative-energy solutions are given by
\begin{equation} \label{xipi_hyperobolicSolutions}
\xi = \xi_{0}\cosh(t - t_{0} ) \,,
\end{equation}
where integration constants are fixed by specifying the turning point $\xi_0$ and the time $t_0$ when the trajectory reaches this turning point,  $\xi(t_0) = \xi_{0}$. The only static solution is $\xi_0 = 0$ and is unstable.  A typical classical phase-space portrait for the IHO with negative energy is drawn in panel (a) of \fref{fig2}. For each negative energy ($E >0$) there are two distinct hyperbolic trajectories depending on whether the particle approaches from the right or left. The trajectory in the left-hand quadrant of the figure describes a particle approaching from the left, while the one in the right-hand quadrant corresponds to a particle approaching from the right.
\par The quantum mechanics of an IHO is well-posed mathematically -- {\it i.e.}~is essentially self-adjoint \cite{reed1975ii, finster2017p, subramanyan2021physics} on the real line---but is unstable because its Hamiltonian is unbounded from below. Parabolic cylinder functions are known to provide an energy eigenbasis, and the invariance of the Hamiltonian under parity ($\xi \to - \xi$) implies each energy level is doubly degenerate. A general solution to the Schr\"odinger equation with energy ${\cal E} = - E = - \hat E \hbar \omega$ can be written 
\begin{eqnarray}\label{pcfs1}
\phi(\xi) &=& C_{1} \, \phi_{1} (\xi)+ C_{2} \,  \phi_{2}(\xi) \nonumber \\
 &=& C_{1} \, D_{{\mathrm{i}\hat E} - \frac12}(\sqrt{2}e^{-\mathrm{i}3\pi/4} \xi) + C_{2} \, D_{-{\mathrm{i}\hat E} - \frac12}(\sqrt{2}e^{-\mathrm{i}\pi/4} \xi)  
\end{eqnarray}
\begin{figure*}
\subfloat[]{\includegraphics[width = 0.45\columnwidth]{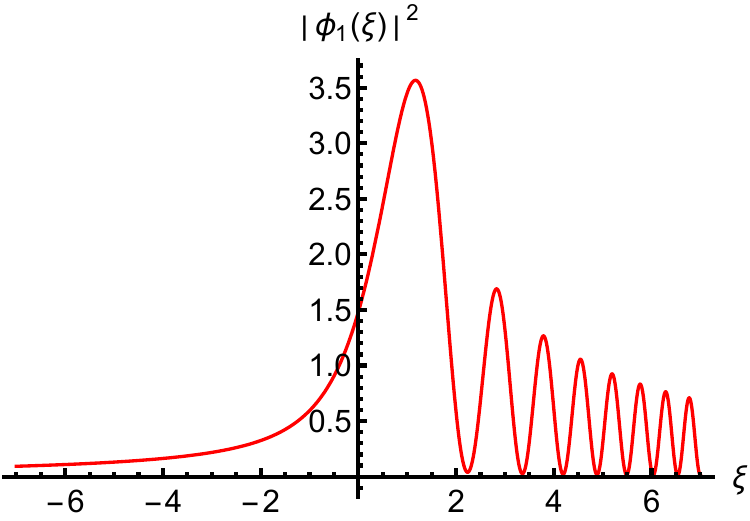}}\hspace{2cm}
\subfloat[]{\includegraphics[width = 0.45\columnwidth]{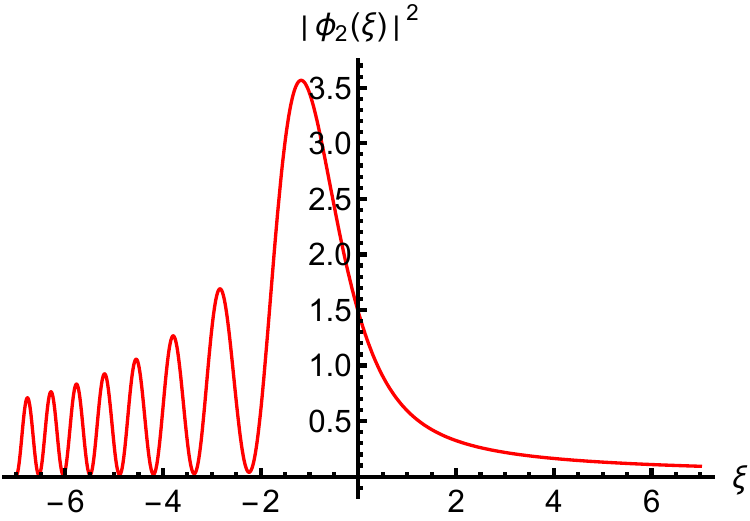}}\hspace{2cm}
\subfloat[]{\includegraphics[width = 0.45\columnwidth]{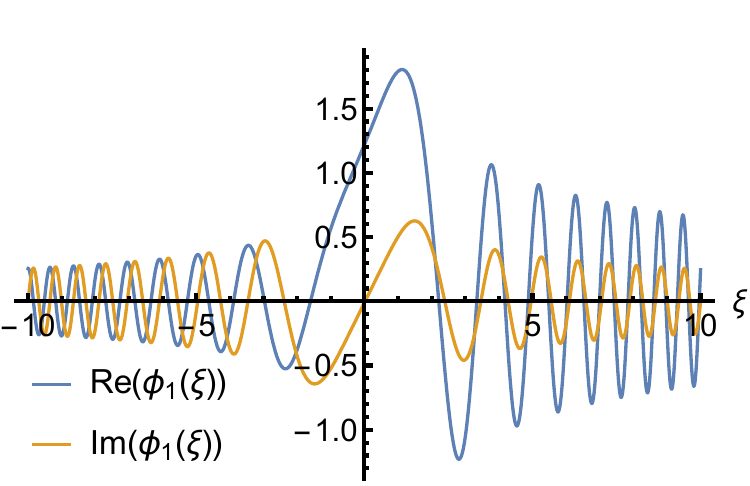}}\hspace{2cm}
\subfloat[]{\includegraphics[width = 0.45\columnwidth]{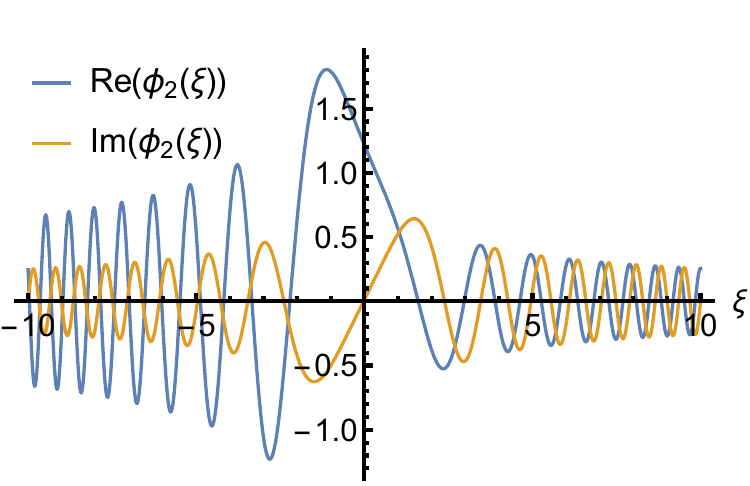}}
\caption{Plots of the two linearly independent solutions of the IHO Schr\"{o}dinger equation. The left two panels show the solution $\phi_{1}(\xi) = D_{\mathrm{i}\hat E -1/2}\left(\sqrt{2} e^{-\mathrm{i}3\pi/4}\xi \right) $ and the right two panels show $\phi_{2} = D_{-\mathrm{i}\hat E -1/2}\left(\sqrt{2} e^{-\mathrm{i}\pi/4}\xi \right)$. Panels (a) and (b) plot the probability density, $\left| \phi(\xi) \right|^{2}$, for each case while panels (c) and (d) show their real and imaginary parts. }
\label{fig3}
\end{figure*}
where $D_{s}(z)$ are parabolic cylinder functions \cite{NIST:DLMF} and $C_{1}, C_{2}$ are arbitrary constants that determine the choice of wavefunction. The eigenfunctions $\phi_{1}$ and $\phi_{2}$ are plotted as a function of $\xi$ in \fref{fig3}. 

Note that although it may seem as though the two eigenfunctions $\phi_{1}$ and $\phi_{2}$ in (\ref{pcfs1}) correspond to different energies due to the $\pm \mathrm{i} \hat{E}$ factors labelling their parabolic cylinder functions, this is actually not the case due the effect of the different complex phases of their arguments. This point is spelled out in \ref{A0}.  Indeed, although we shall not use it, an alternative energy eigenbasis where both terms have the same $+\mathrm{i}\hat{E}$ factors and makes the action of parity more manifest is
\begin{equation}
\tilde{\phi}_{\pm}(\xi) = C_{\pm} D_{\mathrm{i}\hat E - \frac12}(\pm \sqrt{2}e^{-\mathrm{i}3\pi/4} \xi) \,.\label{pcf_parity_conjugates}
\end{equation}   

\par Because the wave-functions are oscillatory at large $\left| \xi \right|$ the states are not normalizable even for negative energies. As is standard for continuum states this means that normalization of the state cannot determine one combination of $C_{1}$ and $C_{2}$. One instead asks scattering questions, such as by specifying an incoming flux at large positive or negative $\xi$ and asking for the transmission and reflection probabilities per unit incident flux. For later purposes we remark that this implicitly means that states are chosen according to boundary information specified at large $\xi$.

\subsection{Berry-Keating system}
\label{BK}

Although not previously emphasized, the IHO Schr\"{o}dinger system is part of a closed $\mbf{su}(1,1)$ Lie algebra \cite{prakash1994mixed,pitaevskii1997breathing,gurappa2003unified,subramanyan2021physics} defined by the generators 
\begin{equation} \label{SU11alg}
  K_{1} = \frac{1}{2}(\pi^{2}-\xi^{2}) \,, \;
  K_{2} = \frac{1}{2}(\pi^{2} + \xi^{2}) \,,\;  
  K_{3} = \frac{1}{2}(\xi \cdot\pi + \pi \cdot\xi) 
\end{equation}
since the commutation relations imply these satisfy
 \begin{equation}\label{su(1,1)algebra}
 \left[K_{1},K_{2}\right] = -\mbf{i}K_{3},\; \left[K_{2},K_{3}\right] = \mbf{i}K_{1}, \; \left[K_{3},K_{1}\right] = \mbf{i}K_{2} \,.
 \end{equation}  
The Casimir invariant $\hat C = K_{3}^{2} - K_{1}^{2} - K_{2}^{2}$ commutes with the IHO Hamiltonian ($K_{1}$), {\it i.e.} $\left[\hat C, K_{1} \right] = 0$,  as well as with the other generators $K_{2}$ and $K_{3}$. This is a spectrum-generating algebra because the generator $K_{1}$ is the IHO Hamiltonian and the other generators do not commute with it. 

This observation suggests a canonical transformation that has the effect of swapping which of the $K_i$ plays the role of Hamiltonian. In particular, we transform to BK variables for which $K_3$ becomes the Hamiltonian. This is done using the following canonical transformation to the new hermitian operators \cite{balazs1990wigner}
\begin{equation}\label{canonical_transf}
   Q = \frac{\pi + \xi}{\sqrt{2}}, \quad P = \frac{\pi - \xi}{\sqrt{2}} \,,
\end{equation}
which preserves the classical Poisson bracket, $\{Q,P\} = 1$, and hence the commutation relation, $\commutator{Q}{P} = \mbf{i}$.  In terms of these the IHO Hamiltonian becomes
\begin{equation}
H(Q,P) = \frac{Q \cdot P + P \cdot Q}{2}  \ .  \label{HamPQ}
\end{equation}
This quantum BK Hamiltonian is symmetric under the exchange of $Q$ and $P$, like the simple harmonic oscillator Hamiltonian. Parity symmetry is realized in these variables as $(Q,P) \to (-Q,-P)$.

\par Hamilton's (classical) equations in the new variables are
\begin{equation}
\frac{\exd Q}{\exd t}  = Q \quad \hbox{and} \quad \frac{\exd P}{\exd t}  = -P
\end{equation}
with solutions \cite{ullinger2022logarithmic}
\begin{equation}\label{time_logQ}
Q = Q_0 \, e^t \quad \hbox{and} \quad P = P_0 \, e^{-t} \,,
\end{equation}
 \begin{figure}
\includegraphics{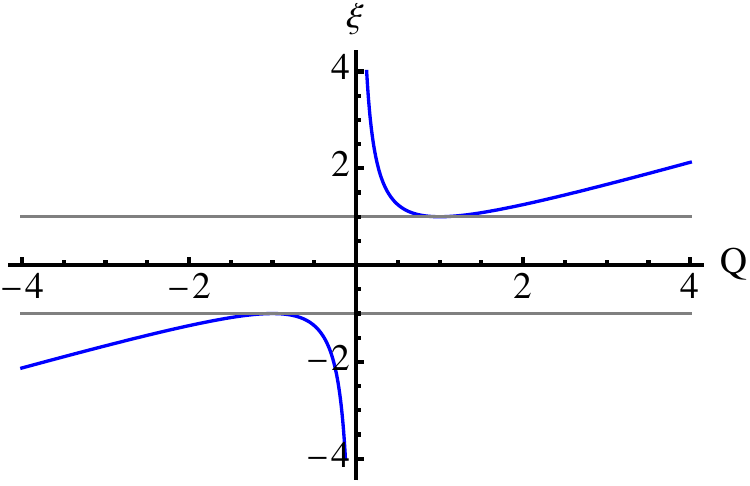}
\caption{The figure shows the relation between the variables $\xi$ and $Q$ given by $\xi = \frac{\xi_{0}}{2}\left(Q + \frac{1}{Q}\right)$. The sectors $Q > 0$ and $Q < 0$ are disconnected classically. Although in general the relationship between $\xi$ and $Q$ is multivalued, we can see that for small $Q$ we have the simple inverse relationship $\xi= \frac{\xi_{0}}{2}\frac{1}{Q}$ so that small distances in the BK and ISP systems are equivalent to large distances in the IHO.}
\label{fig4}
\end{figure} 
where $Q_{0}$ and $P_0$ are the initial conditions at $t=0$ -- see \fref{fig2}. (Notice in particular that $\frac{dQ}{d t}$ is \textit{not} $P$.) These solutions show how time translation corresponds to rescalings of $Q$ and $P$ and so sheds light on the scale-invariance behind the algebra (\ref{SU11alg}). Although the canonical transformation (\ref{canonical_transf}) acts on both position and momentum, it is possible to use the classical solutions to derive a relationship between the position variables $\xi$ and $Q$ alone. Comparing the solutions (\ref{xipi_hyperobolicSolutions}) and (\ref{time_logQ}) leads to the relation
\begin{equation}
 \xi = \frac{\xi_{0}}{2} \left(Q + \frac{1}{Q}\right) \,,
\end{equation}
a plot of which is given in \fref{fig4}. Although the relation between $Q$ and $\xi$ is multivalued, for small $Q$ the relation is simply inverse so that small $Q$ is mapped to large $\xi$.

\par The Schr\"{o}dinger equation in these variables takes the form \cite{brout1995primer, chruscinski2003quantum, subramanyan2021physics, ullinger2022logarithmic}:
\begin{equation}\label{BKHamiltonian_NegativeEnergy}
 \frac{1}{2} \, \left( Q \cdot P + P \cdot Q \right) \ket{\phi} = -\frac{E}{\hbar\omega} \ket{\phi} 
\end{equation}
where we again write the energy eigenvalue as ${\cal E} = - E = - \hat E \hbar \omega$. In the position representation this becomes the following first-order equation for $\phi(Q)$,
\begin{equation} \label{euler_operator_Qrep}
Q\frac{\partial \phi}{\partial Q}  = - \left({\mbf{i} \hat E} + \frac{1}{2}\right)\phi 
\end{equation}
whose scale-invariance under $Q \rightarrow \zeta Q$ is manifest. This reflects the fact that the BK Hamiltonian is itself the generator of scale transformations, since
\begin{equation}
e^{\mbf{i}\zeta H(Q,P)}~Q~e^{-\mbf{i}\zeta H(Q,P)}  =  e^{\zeta}~Q  \quad \hbox{and} \quad e^{\mbf{i}\zeta H(Q,P)}~P~e^{-\mbf{i}\zeta H(Q,P)}  =  e^{-\zeta}~P
\end{equation} 
for constant $\zeta$, as expected from the classical result in (\ref{time_logQ}). 

\par
\begin{figure}[t]
\begin{center}
\graphicspath{{./Plots/}}
\includegraphics[width =0.85\columnwidth]{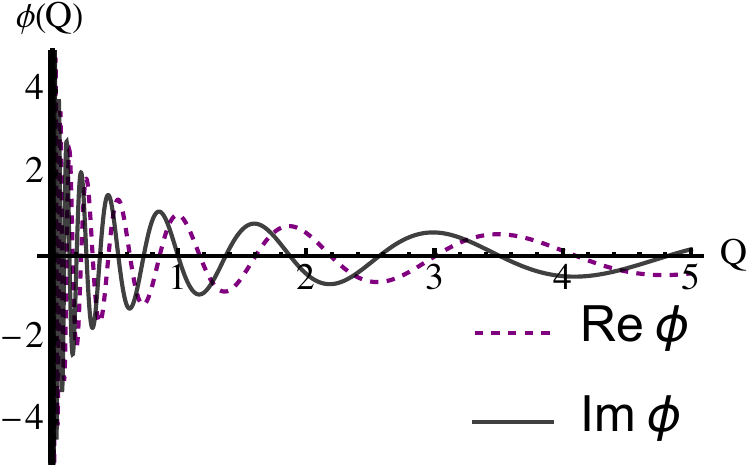}
\end{center}
\caption{An example of an eigenfunction of the BK Hamiltonian $H(Q,P)$ in the Q-representation in $\mathbb{R}^{+}$ given by $Q^{-\mathrm{i}\hat{E}-1/2}=Q^{-1/2}e^{-\mathrm{i} \hat{E} \ln Q}$ with $\hat{E} = 10$. It exhibits both an amplitude singularity and a logarithmic phase singularity at Q = 0.}
\label{fig5}
\end{figure}

 The lone solution of (\ref{euler_operator_Qrep}) is
\begin{equation}\label{BKESol}
\phi(Q) = A~Q^{- \frac12 - \mbf{i}{\hat E}} = A~Q^{-\frac12} e^{- \mbf{i} \hat{E} \ln Q } \, ,
\end{equation}  
where $A$ is a constant. Unusually, because the Schr\"odinger equation is first order there is not a second independent solution to this equation, which at face value seems to contradict the fact that the IHO has doubly degenerate energy eigenspaces. Since parity provides the secondary label for states within a given energy eigenspace we should ask what it implies for the solutions to (\ref{euler_operator_Qrep}). When doing so it is crucial that for general $\hat E$ the solution (\ref{BKESol}) has a logarithmic branch point at $Q = 0$, leading to both an amplitude singularity and a logarithmic phase singularity at $Q = 0 $ [as can be seen in \fref{fig5}]. Such singularities are generic signatures of quantum catastrophes \cite{leonhardt2002theory,kiss2004towards}, and appear in many other interesting physical systems like waves near black-hole event horizons, in accelerated frames, and so on \cite{brout1995primer, coutant2014hawking, subramanyan2021physics, ullinger2022logarithmic, farrell2023logarithmic}. 

\par What is important for present purposes is that the logarithmic branch point at $Q=0$ makes the extrapolation of (\ref{BKESol}) from $Q > 0$ to $Q < 0$ non-unique. Starting with $Q > 0$ and navigating around different sides of the branch point -- such as by multiplying by $e^{\pm \mathrm{i} \pi}$ -- leads to different sheets for $Q < 0$. These two distinct extensions of (\ref{BKESol}) to negative $Q$ provide the two linearly independent energy eigenstates of the BK problem. More formally, the BK Hamiltonian can be shown to be essentially self-adjoint in the half real line ($\mathbb{R}^{\pm}$) but not on the full real line ($\mathbb{R}$) using von Neumann's theorem \cite{melintransform_xp}. 

\par Because the two solutions share the same $Q$-dependence for positive $Q$ (say) their difference is an energy eigenstate that vanishes for all $Q > 0$. A conventional choice for a basis of eigenstates is to ask one basis eigenstate to vanish for $Q > 0$ and the other to vanish for $Q < 0$. This leads to the following expression for the general energy eigenvalue for the BK Hamiltonian \cite{berry1999h}
\begin{equation} \label{Qrepwavefunction}
\phi(Q) = \left| Q\right|^{ - \frac{1}{2} - {\mbf{i}\hat E}} \left[ A\,\Theta(Q) + B\,\Theta(-Q) \right] \,,
\end{equation} 
where $A$ and $B$ are the integration constants and $\Theta(Q)$ is the Heaviside step function. 

\par The existence of a canonical transformation between the IHO and BK systems implies the existence of a canonical map that relates their quantum states \cite{brout1995primer, subramanyan2021physics, ullinger2022logarithmic}. The mapping between the states presented in  (\ref{pcfs1}) and (\ref{Qrepwavefunction}) is provided by a `quantum canonical transform' (see \ref{QCT_BK_first_state})
\begin{equation}\label{pcf_from_BK}
 \phi(\xi) = \int_{0}^{\infty} dQ~Q^{-\frac{1}{2} - \mathrm{i}{\hat E}} e^{\mathrm{i}\left(-\frac{1}{2} \xi^{2} + \sqrt{2}\xi Q - \frac{1}{2}  Q^{2}\right)  }  
\end{equation}
which can be recognized as one of the integral representations of parabolic cylinder functions, as will be discussed in more detail in \Sref{sec:mapping}.

\subsection{Inverse-square potential}

The ISP problem has the Hamiltonian appropriate to an interaction potential whose strength falls off like the square of the distance from the origin
\be
   H = \frac{1}{2} \, P^2  - \frac{g}{Q^2} \,,
\ee
and our focus here is on attractive potentials for which $g > 0$. This Hamiltonian also enjoys a spectrum-generating scale invariance because $H \to \zeta^{-2} H$ under the scale transformation $Q \rightarrow \zeta Q$ \cite{case1950singular}. Classically,  the generator of the scale transformation is $D = Q P$, which satisfies
\begin{equation} \label{Drate}
\frac{\exd D}{\exd t} = 2H
\end{equation}  
and so is not conserved unless restricted to configurations with vanishing energy. Eq.~\pref{Drate} is called an \textit{almost} conservation law \cite{rajeev2013advanced}. 

\par The quantum version of this problem has the time-independent Schr\"odinger equation
\be \label{SchEgk}
Q^2 \frac{\partial^2 \psi}{\partial Q^2}  + \left( 2g - \kappa^2 Q^2 \right) \psi = 0\,,
\ee
for energy eigenvalue ${\cal E} = -E = - \frac{1}{2} \kappa^2$. The change of variables $\psi(z) = z^l \, e^{-z/2} u(z)$ for $z = 2\kappa Q$ and $l$ satisfying $l = \frac{1}{2} ( 1 + \zeta)$ with 
\be
\zeta := \sqrt{1 - 8g}  \,,
\ee
leads to a new dependent variable $u(z)$ that satisfies the confluent hypergeometric equation. Notice that $\zeta$ changes from real to imaginary at what is called the critical coupling $g_c = \frac18$.

\par The two linearly independent solutions are 
\begin{equation} \label{InvSqWavefunction}
	\psi_\pm(Q) = (2\kappa Q)^{\frac 12\left(1 \pm\, \zeta\right)}e^{-\kappa Q}M\left[\frac 12\left(1 \pm \zeta\right), 1 \pm \zeta; 2\kappa Q\right] \,,
\end{equation}
provided $1\pm \zeta$ is not a nonpositive integer, where $M(a,b;z) = 1 + (a/b) z + \cdots$ is the confluent hypergeometric function.

\par Famously, when $g > 0$ {\it both} of the solutions (\ref{InvSqWavefunction}) can be singular at $Q = 0$ and so (unlike for the Coulomb problem) one cannot use the regularity of the solution at the origin as a criterion for selecting one or the other. Instead the eigenvalue problem for this Hamiltonian is not well-posed without specifying a boundary condition \cite{case1950singular} at $Q = 0$. Physically this conveys how the solutions depend on the properties of whatever the object is that sits at $Q = 0$ (and so is ultimately responsible for the existence of the $1/Q^2$ potential). Although this boundary condition is often chosen to ensure the Hamiltonian is self-adjoint, this need not be what is required by specific physical situations (such as when the origin is a source or sink of probability \cite{plestid2018fall}). 

\par In practice the divergence of solutions at $Q=0$ usually means any boundary condition is actually imposed at a small but nonzero $Q = \epsilon$. This boundary condition is often linear (Robin-type boundary condition) and when it is it can be written
\be \label{RobinBC}
   \left.  \frac{\partial \psi}{\partial Q} \right|_{Q=\epsilon} = \lambda \, \psi(\epsilon) \,,
\ee
for some constant $\lambda$. Conditions of this form are sometimes also referred to as Bethe-Peierls boundary conditions following their early application in nuclear physics \cite{bethe1935quantum}.  A modern systematic method for determining boundary conditions at small distances is provided by PPEFT wherein the boundary condition can be related to an effective action describing the object at the origin and because of this dimensional arguments can be applied that typically lead to (\ref{RobinBC}) at low energy \cite{cliffppeft1}.  In the present context of the scale invariant ISP, imposing a boundary condition at nonzero $Q$ breaks scale invariance and this ultimately causes anomaly-type quantum breaking of the classical scale invariance. 
Furthermore, because the regularization scale $\epsilon$ is arbitrary it cannot appear in physical predictions. This turns out to be ensured by an implicit $\epsilon$-dependence carried by the parameter $\lambda$, which adjusts as a function of $\epsilon$ in a way that keeps physical observables fixed; an adjustment that is captured by a renormalization-group flow $\lambda = \lambda(\epsilon)$ \cite{cliffppeft1}. Comparison with well-understood systems (such as atoms) with small objects at the origin (nuclei) shows that the physical scale associated with the physics at $Q=0$ is ultimately a renormalization-group invariant of this flow \cite{cliffppeft2, cliffppeft3, cliffppeftHe, cliffppeftH}. 

\section{Mapping between the inverted harmonic oscillator and the inverse square potential system}
\label{sec:mapping}

We now construct the mapping between the IHO/BK system and the ISP system. 

\subsection{Mapping of the Hamiltonian}
\label{squared_BK}

The main construction in the duality mapping relates the {\it square} of the BK Hamiltonian to the ISP Hamiltonian. The eigenstates $\phi(Q)$ of the BK Hamiltonian are also eigenstates of its square but with a squared eigenvalue
%
\begin{equation}\label{QP_square}
\left(\frac{Q\cdot P + P\cdot Q}{2} \right)^{2} \phi(Q) =  \hat E^{2} \phi(Q) \ .
\end{equation}  
 In the position representation this equation for the squared BK Hamiltonian can easily be seen to take the form
\begin{equation}\label{QP_squareQ}
\left[ Q^{2}\frac{\partial^{2}}{\partial Q^{2}} + 2\,Q\frac{\partial }{\partial Q} + \Bigl(\hat{E}^{2} + \frac14 \Bigr) \right] \phi(Q) = 0  \,.
\end{equation}
Using the integrating factor 
\be\label{BKtoInvSqPsi}
  \phi(Q) = \frac{\chi(Q)}{Q} \,,
\ee
then gives
\begin{equation}\label{inverse_sq_eqn}
  -\frac{\partial^{2}\chi(Q)}{\partial Q^{2}} - \frac{\left( \hat{E}^{2}+\frac14\right)}{Q^{2}}\chi(Q) = 0 \ ,
\end{equation} 
which with the definition
\be \label{Evsg}
   2g = \hat{E}^{2}+\frac14  \geq \frac14 \,,
\ee
is the ISP Schr\"odinger equation (\ref{SchEgk}) restricted to the case of zero energy ($\kappa^2 = 0$). Notice that the restriction to $\kappa^2 = 0$ ensures that the condition (\ref{Drate}) becomes an honest-to-God conservation rule. The condition $2g > \frac14$ means $g > g_c = \frac18$ and so the coupling given by our mapping is always super-critical. 

\par The solution $\chi(Q)$ for the ISP Schr\"{o}dinger equation specialized to zero energy is particularly simple because the confluent hypergeometric equation degenerates to the Euler equation, which has power-law solutions. The most general solution is
\begin{equation} \label{wavefunction_inverse_square}
\chi(Q) = \alpha Q^{\frac{1}{2} - {\mathrm{i}\hat E}} + \beta Q^{\frac{1}{2} + {\mathrm{i}\hat E}} 
\end{equation} 
where $\alpha$ and $\beta$ are integration constants. These two basis solutions are linearly independent if $\hat E \neq 0$ (the second solution involves $\log Q$ if $\hat E \neq 0$). They also exhibit logarithmic phase singularities as did Eq.~(\ref{Qrepwavefunction}) for the BK Hamiltonian. We see that the effect of the squaring is to allow $\pm \hat{E}$ states (these do \textit{not} correspond to energy on the ISP side of the mapping).

\par From this point of view the boundary-condition ambiguity at the origin in the ISP system maps directly onto the boundary condition ambiguity of the IHO problem; in both cases normalizability cannot be used to choose a preferred state and a boundary condition is instead required near a singular point (at $Q$ near zero for the ISP problem and at large $\xi$ for the IHO problem). This implies in particular that the entire renormalization-group description for the ISP problem \cite{cliffppeft1, plestid2018fall} can be directly mapped across to the IHO (and its applications).   

\subsection{Mapping the states}

Equation (\ref{BKtoInvSqPsi}) shows explicitly how zero-energy states $\chi(Q)$ for an attractive ISP with super-critical coupling $g$ are mapped to energy eigenstates $\phi(Q)$ of the BK system with energy $E(g)$ given by (\ref{Evsg}). The quantum canonical transformation described in (\ref{pcf_from_BK}) then maps the result onto an IHO state $\phi(\xi)$. 

\par In this language the  zero-energy ISP solution
\be 
 \chi(Q) = \alpha \, Q^{\frac{1}{2} - \mathrm{i}\hat{E}} 
\ee
becomes the BK solution
\be
  \phi(Q) =  \alpha \, Q^{-\frac{1}{2} - \mathrm{i}\hat{E}}  
\ee
and this in turn maps over to the following IHO eigenstate via \cite{brout1995primer} 
 \begin{equation} \label{pcf_from_BK2}
 \phi_{1}(\xi) = \alpha\int_{0}^{\infty} \exd Q~ Q^{-\mathrm{i}\hat E - \frac{1}{2}} e^{\mathrm{i}\left(-\frac{1}{2}\xi^{2} + \sqrt{2}\xi Q - \frac{1}{2}Q^{2} \right)  } \,.
 \end{equation}
Noting that parabolic cylinder functions $D_{s} (z)$ have the integral representation  \cite{ullinger2022logarithmic}
\begin{equation}\label{integral_rep_Pcf}
 D_{s} \left(\frac{b}{\sqrt{2 a}} \right) = \frac{\left(2a\right)^{-s/2}}{\Gamma(-s)} \int_{0}^{\infty} \exd Q~ Q^{-s-1} e^{-aQ^{2} - bQ - \frac{b^{2}}{8a}} 
 \end{equation}
and putting $a = \mathrm{i}/2$, $s =\mathrm{i}\hat{E} - \frac{1}{2}$, $b = -\sqrt{2}\mathrm{i}\xi$, the integral in (\ref{pcf_from_BK2}) becomes 
%
\begin{equation}
\phi_{1}(\xi) = \alpha \,  e^{-\frac{\pi \hat E}{4} } e^{-\mathrm{i}\pi/8} \,  \Gamma\left(\frac{1}{2} - {\mathrm{i}\hat E}\right) \, 
 D_{{\mathrm{i}\hat E} - \frac12} \left( \sqrt{2} e^{-\mathrm{i}3\pi/4} \xi  \right) \label{first_pcf_from_BK_explicit} 
 \end{equation}
which is the first of the IHO wavefunctions. Its coefficient $\alpha$ maps over directly from the first ISP solution since the mapping does not mix in any of the second IHO wavefunction $\phi_{2}(\xi)$.
 A similar construction applies to the second ISP solution $\beta \, Q^{{\mathrm{i}\hat E} - \frac{1}{2}}$, which maps across to the IHO state
 \begin{equation}\label{second_pcf_from_BK}
 \phi_{2}(\xi) =\beta \int_{0}^{\infty}\exd Q~ Q^{{\mathrm{i}\hat E} - \frac12} e^{\mathrm{i}\left(\frac{1}{2}\xi^{2} + \sqrt{2}\xi Q + \frac{1}{2}Q^{2} \right)  } \,.
 \end{equation}
See \ref{second_state_details} for details on how to obtain the above form of kernel that generates the second IHO state.  Using (\ref{integral_rep_Pcf}) again, but with $a = -\frac{\mathrm{i}}{2}$, $b = -\sqrt{2}\mathrm{i}\xi$ and $-s =\frac12 + \mathrm{i}{\hat E}$, allows the second solution on the IHO side to be written
%
\begin{equation}
\phi_{2}(\xi) =  \beta e^{- \frac{\pi\hat E}{4}} e^{\mathrm{i}\pi/8} \, \Gamma\left(\frac{1}{2} + \mathrm{i}\hat{E} \right) \, D_{-{\mathrm{i}\hat E} - \frac{1}{2}}\left(\sqrt{2}e^{-\mathrm{i}\pi/4}\xi\right) \,.  \label{second_pcf_from_BK_explicit}  
\end{equation}

\par It follows that the general zero-energy ISP eigenstate [$\chi(Q) = \alpha Q^{\frac{1}{2} - {\mathrm{i}\hat E}} + \beta Q^{\frac{1}{2} + {\mathrm{i}\hat E}} $] given in (\ref{wavefunction_inverse_square})
maps over to the IHO state [$\phi(\xi) = C_{1} D_{{\mathrm{i}\hat E} - \frac{1}{2}}(\sqrt{2}e^{-\mathrm{i}3\pi/4} \xi) + C_{2} D_{-{\mathrm{i}\hat E} - \frac{1}{2}}(\sqrt{2}e^{-\mathrm{i}\pi/4} \xi)$] given in (\ref{pcfs1})
%
where the constants $C_1$ and $C_2$ are expressed in terms of $\alpha$ and $\beta$ by
\begin{equation}
  C_1 = \alpha \,  e^{-\frac{\pi\hat E }{4}} e^{-\mathrm{i}\pi/8} \Gamma\left(\frac{1}{2} - {\mathrm{i}\hat E}\right)  \quad , \quad
 C_2 = \beta \, e^{-\frac{\pi\hat E}{4}} e^{\mathrm{i}\pi/8} \Gamma\left(\frac{1}{2} +{\mathrm{i}\hat E} \right) \,.
\end{equation}
Physical predictions depend only on the ratios $\alpha/\beta$ and $C_1/C_2$ and so are related by 
\be \label{ratiovsratio}
  \frac{C_1}{C_2} = \left( \frac{\alpha}{\beta} \right)  \frac{ \Gamma\left(\frac{1}{2} - { \mathrm{i}\hat E} \right) }{   \Gamma\left(\frac{1}{2} +{\mathrm{i}\hat E} \right)}\;  e^{-\mathrm{i}\pi/4}\,.
\ee

\subsection{Mapping the boundary condition}
 
The previous sections show in detail how the zero-energy states of the super-critical ISP system are mapped onto the states of the IHO system. Expression (\ref{ratiovsratio}) is the core of a complete solution to the question of how to map observables (like scattering rates) in the IHO onto similar observables in the ISP system. Once the observable is known as a function of $\alpha/\beta$ or $C_1/C_2$ then this map can be used to relate observables directly to one another.

It can be more useful to directly relate the boundary condition that makes the ISP well-defined with the boundary condition used for the IHO. That is, suppose the ISP imposes the linear boundary condition (\ref{RobinBC}) with constant $\lambda_{\rm ISP}$ at $Q=\epsilon$ and the IHO involves a similar boundary condition at large $\xi = L$ (with constant $\lambda_{\rm IHO}$). How are the parameters $\lambda_{\rm ISP}$ and $\lambda_{\rm IHO}$ related by the map between these two systems? Finding this relation is our purpose in the present section. We do so by using \pref{ratiovsratio} together with the predictions for $\alpha/\beta$ as a function of the pair $(\lambda_{\rm ISP}, \epsilon)$ and for $C_1/C_2$ as function of $(\lambda_{\rm IHO},L)$.

An important complication arises because the mapping from the pair $(\lambda,\epsilon)$ to a ratio like $C_1/C_2$ is many to one. It is many to one because the scale $\epsilon$ is arbitrary and so $\lambda$ necessarily varies as $\epsilon$ does, in precisely the way required to ensure that observables (and so also ratios like $\alpha/\beta$ and $C_1/C_2$) remain $\epsilon$-independent. Because of this the prediction for observables from boundary conditions proceeds in two steps. First one identifies the RG-invariant quantities that characterize the curve $\lambda(\epsilon)$. Second a formula is derived for $\alpha/\beta$ or $C_1/C_2$ as a function of these RG invariants (see \cite{cliffppeft1} for details). We therefore pause briefly to recap how $\lambda_{\rm ISP}$ and $\lambda_{\rm IHO}$ run.

\medskip\noindent{\bf ISP Case:}
We start by reviewing the zero-energy ISP case \cite{cliffppeft1}, for which the energy eigenstates are
\begin{equation} \label{wavefunction_inverse_square2}
\chi(Q) = \alpha Q^{\frac{1}{2} - \mathrm{i}\hat E} + \beta Q^{\frac{1}{2} + \mathrm{i}\hat E } \,,
\end{equation} 
where $\hat E := E/\hbar\omega$ is computed from the ISP coupling $g$ using (\ref{Evsg}). We wish to determine the ratio $\alpha/\beta$ that follows from (\ref{RobinBC}), which we rewrite in the equivalent form
\be \label{RobinBC2}
   \lambda_{\rm ISP} = \frac{1}{\chi (\epsilon)} \left( \frac{\partial \chi}{\partial Q} \right)_{Q=\epsilon}  \,.
\ee

The relation between $\lambda_{\rm ISP}$ and $\alpha/\beta$ is found by substituting in (\ref{wavefunction_inverse_square2}) for $\chi(Q)$. The result is
\begin{equation} \label{LambdavsConsts}
\frac{{\Lambda}_{\rm ISP}}{\mathrm{i}\hat E} =  \frac{1 - (\alpha/\beta)~\epsilon^{-2\mathrm{i}\hat E} }{1 + (\alpha/\beta)~\epsilon^{-2\mathrm{i}\hat E} }  \, , 
\end{equation}
where
\be \label{LambdaisDef}
\Lambda_{\rm ISP} := \epsilon \lambda_{\rm ISP}  - \frac12 \,.
\ee
There are two ways to read these last two expressions. First they can be solved for $\alpha/\beta$, giving the solution for the integration constants as a function of the boundary data $(\lambda,\epsilon)$
\be \label{alphabetavsLambda}
   \frac{\alpha}{\beta} = \frac{1 + \mathrm{i} (\Lambda_{\rm ISP}/\hat E)}{1 - \mathrm{i}(\Lambda_{\rm ISP}/\hat E)}   \; \epsilon^{2 \mathrm{i} \hat E} \,.
\ee
The second way to read \pref{LambdavsConsts} and \pref{LambdaisDef} is as a solution to the question of how $\lambda_{\rm ISP}$ must vary as a function of $\epsilon$ if changes to $\epsilon$ are to not change $\alpha/\beta$ (which controls the size of observables). This defines a renormalization group running to the extent that it dictates how $\lambda$ must depend on $\epsilon$ in order for the precise value of $\epsilon$ not to matter for physical predictions. In this view the $\epsilon$-dependence of \pref{alphabetavsLambda} is telling us that physics depends only on the curve $\{\lambda(\epsilon),\epsilon\}$, and so $\alpha/\beta$ depends only on the properties that specify this curve---perhaps an initial condition $\lambda(\epsilon_0) = \lambda_0$, though any other RG-invariant characterization works equally well. In fact, for the ISP system it is known that RG flow in the super-critical regime gives rise to curves $\lambda(\epsilon)$ that are generically limit cycles which encircle but never reach one of two fixed points in the complex $\lambda$-plane (the fixed points are complex conjugates of each other) \cite{bawin2003invsqrg,mueller2004renormalization,kaplan09,moroz2010nonrelativistic}. Complex $\lambda$ indicates that the theory is not self-adjoint and can describe the absorption or emission of particles at the origin. In fact, the behaviour of the fixed points of the boundary condition in the super-critical ISP case is an example of a $\mathcal{PT}$ symmetry breaking transition \cite{sriram_symmetry} that is more commonly studied in the eigenvectors of non-hermitian quantum mechanical systems \cite{Bender_2007}. The flow around a limit cycle gives rise to log-periodic behaviour of physical observables (such as elastic and non-elastic scattering cross-sections) as a functions of experimentally tuneable parameters like incident energy \cite{cliffppeft1,plestid2018fall}.

A convenient choice for the RG-invariant characterization of $\Lambda_{\rm ISP}(\epsilon)$ is given by the pair $(\epsilon_\star, y_\star)$, where $\epsilon = \epsilon_\star$ is the scale where the trajectory $\Lambda_{\rm ISP}(\epsilon)$ crosses the imaginary axis, taking the value $\Lambda_{\rm ISP}(\epsilon_*) = \mathrm{i} y_\star$. This is convenient because the RG-invariant scale $\epsilon_\star$ corresponds to the physical scattering length once $\alpha/\beta$ is computed and converted into a scattering cross section.
The differential version of the running is easier if \pref{alphabetavsLambda} is differentiated holding $\alpha/\beta$ and $\hat E$ fixed, and implies
\be \label{RGIScase}
   \epsilon \, \frac{\exd}{\exd \epsilon} \left( \frac{ \Lambda_{\rm ISP}}{\mathrm{i}\hat E}\right) = \mathrm{i}\hat E \left[ 1 - \left( \frac{\Lambda_{\rm ISP}}{\mathrm{i} \hat E} \right)^2 \right] \,.
\ee 
This shows how the pure imaginary choices $\Lambda_{\rm ISP} = \pm \mathrm{i} \hat E$ are the only fixed points. Using these in \pref{alphabetavsLambda} shows that these fixed points correspond to boundary conditions that set either $\alpha$ or $\beta$ to zero, corresponding to purely incoming or outgoing waves \cite{plestid2018fall, burgess2018effective}.  Notice also that the flow \pref{RGIScase} maps the real axis to itself and so preserves the reality of $\Lambda_{\rm ISP}$ in the special case where the initial condition is real.  

\medskip\noindent{\bf IHO Case:}

A similar story goes through for the  IHO. For large $\xi$ the asymptotics of the parabolic cylinder functions imply the energy eigenstates of the IHO are given by \cite{NIST:DLMF}
\begin{equation} \label{wavefucntion_IHO_asymptotic}
\phi(\xi)  \sim \frac{C_1}{\sqrt{\xi}} e^{\mbf{i}\left(\frac{1}{2}\xi^{2} - \hat{E} \ln(\sqrt{2}\xi) + \frac{1}{2}\theta + \frac{\pi}{4}\right)}  +  \frac{C_2}{\sqrt{\xi}} e^{-\mbf{i}\left(\frac{1}{2}\xi^{2} - \hat{E} \ln(\sqrt{2}\xi)+ \frac{1}{2}\theta + \frac{\pi}{4}\right)} 
\end{equation}
%
where $C_1$ and $C_2$ are the integration constants in \pref{pcfs1}. Here $\theta = \mathrm{arg}~\Gamma (\frac12 + \mathrm{i}\hat E) $. 

The choices $C_1 = 0$ or $C_2 = 0$ correspond to waves asymptotically propagating only in one direction, as can be seen by combining \pref{wavefucntion_IHO_asymptotic} with the time-dependence $e^{-\mathrm{i} {\cal E} t/\hbar} = e^{+\mathrm{i} Et/\hbar}$ and noting that the direction of propagation for the wavefunction can be evaluated using the group velocity  obtained from the total phase of the wavefunction $\Phi(\xi)=\mathrm{Arg}[\phi(\xi)]$ \cite{barton1986quantum}. If the definition of the local wavenumber is defined to be $K(\xi) = \frac{\partial \Phi}{\partial \xi}$ then the definition of the group velocity is $v_{g} = \left(\frac{\partial K}{\partial E}\right)^{-1}$.  

\begin{figure}[t]
\graphicspath{{./Plots/}}
\includegraphics{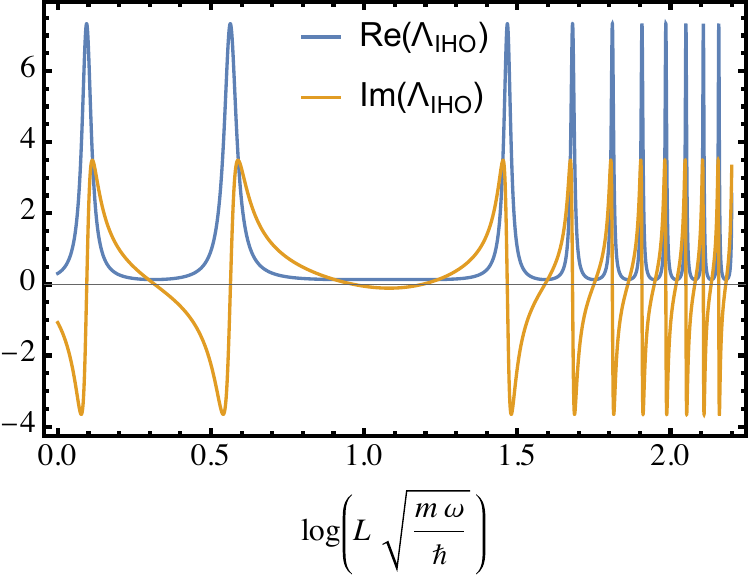}
\caption{The RG flow of the real and imaginary parts of the IHO boundary condition $\Lambda_{\mathrm{IHO}}$ as a function of the logarithm of the scale $L$ at which the boundary condition is applied. The flow is obtained as solutions to the RG evolution equation Eq.~(\ref{betafunction_IHO})  for a particular initial condition $\Lambda_{0} = 7 + \mathrm{i} 1.5 $ and $\hat E = 8.7$. For ISP systems the RG flow is log-periodic, meaning that the period is constant as a function of $\log [L]$, whereas for the generic IHO case shown here the RG flow is chirped because the frequency increases.}
\label{fig6}
\end{figure} 

\begin{figure}[t]
\graphicspath{{./Plots/}}
\includegraphics{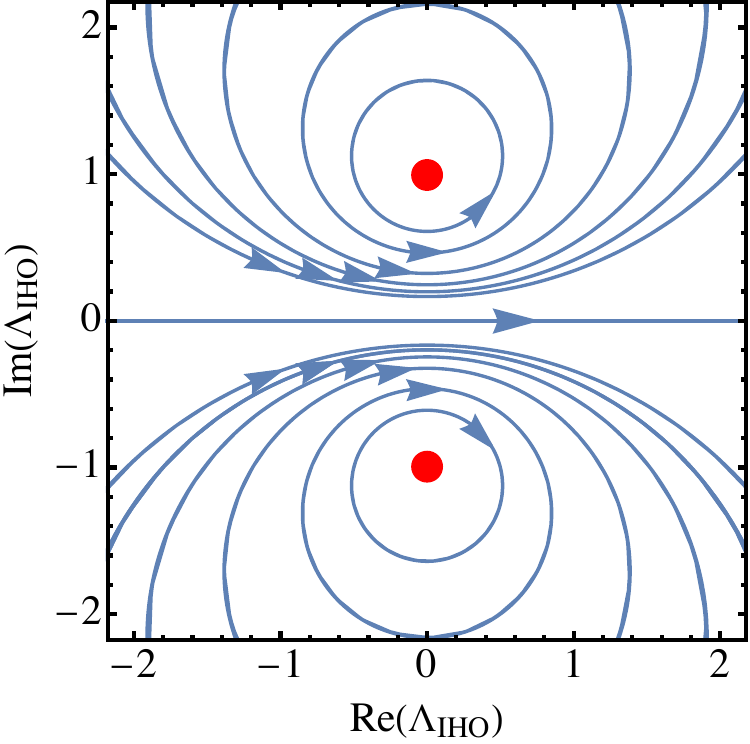}
\caption{The RG flow for the IHO found by solving the differential equation given in Eq.~(\ref{betafunction_IHO})---an analytical solution is given in \ref{A2}. We see that the boundary condition $\Lambda_{\rm IHO}$ exhibits limit cycles in the complex plane as the scale $L$ is changed, where the arrows indicate direction of flow as $L$ increases. Each curve corresponds to a single physical situation, \textit{i.e.} a fixed ratio of the coefficients $C_{1}/C_{2}$. The fixed points are purely imaginary and form a complex conjugate pair and are shown as red dots in the figure. The physics associated with complex values of $\Lambda_{\rm IHO}$ is non-hermitian, and only in the special case where $\Lambda_{\rm IHO}$ starts on the real axis does it remain real during the flow. }
\label{fig7}
\end{figure} 

If the ratio $C_1/C_2$ is determined by a boundary condition of the type \pref{RobinBC} then using \pref{wavefucntion_IHO_asymptotic} in $\lambda_{\rm IHO} = \phi^{-1} (\partial \phi/\partial \xi)_{\xi = L}$ gives an explicit relation 
\begin{equation} \label{LambdaihoCC}
\Lambda_{\rm IHO} = \mathrm{i}\frac{ 1 - ({C_1}/{C_2} ) e^{2\mbf{i}\Omega(L)} }{1 + ({C_1}/{C_2}) e^{2\mbf{i}\Omega(L)}  } \, ,
\end{equation}
where the dimensionless quantity $\Lambda_{\rm IHO}$ is a rescaled version of $\lambda_{\rm IHO}$ and is defined to be
\begin{equation}
\Lambda_{\rm IHO} := \frac{\sqrt{m\omega/\hbar }\, L \lambda_{\rm IHO} + 1/2}{\left[\hat E -( { m\omega L^2}/{\hbar})  \right] } \, ,
\end{equation} 
and the phase $\Omega (L)$ is defined as
\begin{equation}
\Omega(L) :=  \frac{m\omega L^{2}}{2\hbar} - \hat{E}\ln\left(\sqrt{\frac{2m\omega}{\hbar}}L\right) + \frac{\theta}{2} + \frac\pi4 \, .
\end{equation}
Similarly to the ISP case, these expressions can be read as defining how $\lambda_{\rm IHO}$ must depend on $L$ in order to ensure that $C_1/C_2$ is $L$-independent, as well as giving an explicit formula for $C_1/C_2$ as a function of the curve $\lambda_{\rm IHO}(L)$. 

%

The differential evolution of $\Lambda_{\rm IHO}$ can be written as
\begin{equation}\label{betafunction_IHO}
L\frac{\exd \Lambda_{\rm IHO}}{\exd L} = \left( \frac{m\omega L^{2}}{\hbar} - \hat E \right) \left(\Lambda_{\rm IHO}^{2} + 1 \right)
\end{equation}
which reveals fixed points at $\Lambda_{\rm IHO} = \pm \mathrm{i}$. A sample solution to this equation (the analytical solution is given in \ref{A2}) is shown in \fref{fig6} where we see that the RG flow of the real and imaginary parts of $\Lambda_{\mathrm{IHO}}$ as the length scale $L$ is varied is reminiscent of the log-periodic behaviour of the ISP problem, compare with figure 2 in reference \cite{plestid2018fall}. Log-periodic behaviour indicates that the continuous scaling symmetry of the classical problem is broken down to a discrete scaling symmetry by a quantum anomaly. The trajectories in the complex $\Lambda_{\rm IHO}$ plane are limit cycles as shown in \fref{fig7}, {\it i.e.} their geometry in the complex plane is just like that in the ISP. However, the flow along these limit cycles is \textit{faster} than log periodic; the frequency gradually increases as $\log(L)$ increases giving rise to a chirped behaviour as seen in \fref{fig6}. 

Whereas continuous scaling symmetry is explicit in the hamiltonian of the ISP problem, {\it i.e.} the hamiltonian commutes with the scaling operator, the IHO system does not appear to have scaling symmetry. The question is then what continuous symmetry is being broken to give rise to the limit cycles in its RG flow? The answer is that the IHO hamiltonian is part of a larger symmetry group, namely the SU(1,1) group associated with the spectrum generating algebra discussed in \Sref{BK} [Eq.~(\ref{su(1,1)algebra})]. Following reference \cite{pitaevskii1997breathing}, we therefore refer to this as a \textit{hidden} symmetry of the IHO and the limit cycles indicate an anomalous breaking of this hidden symmetry by the linear boundary condition imposed at long distances. 

Clearly the map \pref{ratiovsratio} between $\alpha/\beta$ and $C_1/C_2$ together with expressions like \pref{LambdavsConsts} and \pref{LambdaihoCC} allow boundary condition parameters like $\lambda_{\rm ISP}$ and $\lambda_{\rm IHO}$ to be related to one another. This is most usefully done by relating the RG-invariant characterization of these couplings on either side of the mapping. 

\section{Conclusions}
\label{sec:conclusions}

It has been known for some time that the IHO problem is equivalent to the BK problem through an explicit canonical transformation. In this paper we provide a precise mapping between this joint system and a particle in an ISP with a super-critical attractive coupling. The map relates the zero-energy subspace of the ISP problem to the eigenstates of the IHO/BK system, with the IHO energy ${\cal E} = - E = - \hat E \hbar \omega$ being mapped into the strength of the ISP $-g/Q^2$ through expression \pref{Evsg}.
 
Besides showing how the Hamiltonians of these systems are related we also explicitly identify how a convenient basis of energy eigenstates are mapped into one another. Physical applications in both systems rely on the specification of boundary conditions and we use our knowledge of how the states are mapped to see how the boundary conditions are also related [in the typical case where the boundary condition is linear, as defined in equation (\ref{RobinBC})]. 

Having these systems be explicitly related explains why they share distinctive features such as classical scale invariance with quantum anomalies. Many systems reduce to these models in particular limits ({\it e.g.}~Schwinger pair-production can be related to solutions of the IHO problem) and one hopes the mapping described here will help find new connections amongst these ancilliary systems.   

 \ack
CB and DO acknowledge funding from the Natural Sciences and Engineering Research Council of Canada (NSERC) through Discovery Grants. Research at the Perimeter Institute is supported in part by the Government of Canada through the Department of Innovation, Science and Economic Development and by the Province of Ontario through the Ministry of
Colleges and Universities. SS dedicates this paper to the memory of his beloved mother Chithra Narayanan.

 \appendix
 
 \section{Verification that the two solutions $\phi_{1}(\xi)$ and $\phi_{2}(\xi)$ of the IHO have the same energy}
 \label{A0}
In \Sref{fall_to_infinity} of the main body of the paper we started off with an inverted harmonic oscillator Schr\"{o}dinger equation in the form
 \begin{equation}
 \frac{\pi^{2} - \xi^{2}}{2} \phi(\xi) = -\hat E\phi(\xi) \ .
 \end{equation}
In the $\xi$ representation this can be written as 
 \begin{equation}
\left( -\frac{1}{2}\frac{d^{2}}{d\xi^{2}} - \frac{1}{2} \xi^{2}\right)\phi = -\hat E \phi \label{IHO_equation}  \ .
 \end{equation}
 We shall now check if both the solutions $\phi_{1}(\xi)$ and $\phi_{2}(\xi)$ of the IHO given in Eq.~(\ref{pcfs1}) are states with the same energy as claimed. 
  \begin{enumerate}
  \item[(i)] We first note that the standard parabolic cylinder differential equation that is satisfied by $\phi = D_{\nu}(z)$ is given in \cite{NIST:DLMF} as
 \begin{equation}
 \frac{d^{2}\phi}{d z^{2}} + \left(\nu + \frac{1}{2} - \frac{z^{2}}{4} \right)\phi = 0 \ . \label{pcfD_differential_eqn}
 \end{equation}
\item[(ii)] To find the differential equation whose solution is $\phi_{1}(\xi) = D_{\mathrm{i}\hat E - \frac{1}{2}}\left(\sqrt{2}e^{-\mathrm{i}3\pi/4} \xi\right)$, we put
\begin{equation}
z = \sqrt{2}e^{-\mathrm{i}3\pi/4} \xi
\end{equation}
which means that $\frac{z^{2}}{4} = \frac{\mathrm{i}}{2} \xi^{2}$.
We also write the index as $\nu = \mathrm{i}\hat E - \frac{1}{2}$. The parabolic cylinder differential equation in Eq.~(\ref{pcfD_differential_eqn}) then becomes
\begin{equation}
\left(-\frac{\mathrm{i}}{2}\frac{d^{2}}{d\xi^{2}} + \mathrm{i}\hat E - \frac{\mathrm{i}}{2}\xi^{2} \right) \phi_{1} = 0 
\end{equation}
which can be written as 
\begin{equation}
\left(-\frac{1}{2}\frac{d^{2}}{d\xi^{2}} - \frac{1}{2} \xi^{2}\right)\phi_{1} = -\hat E \phi_{1}
\end{equation}
which is same as Eq.~(\ref{IHO_equation}).
\item[(iii)] Next we find the underlying differential equation for the second solution $\phi_{2}(\xi) = D_{-\mathrm{i}\hat E - \frac{1}{2}}\left(\sqrt{2}e^{-\mathrm{i}\pi/4}\xi\right)$. This time we put
 \begin{equation}
 z = \sqrt{2} e^{-\mathrm{i}\pi/4} \xi 
 \end{equation}
 which implies that $\frac{z^{2}}{4} = -\frac{\mathrm{i}}{2} \xi^{2}$.
Writing the index as $\nu = -\mathrm{i}\hat E - \frac{1}{2}$ the same parabolic cylinder equation Eq.~(\ref{pcfD_differential_eqn}) now becomes
 \begin{equation}
 \left(\frac{\mathrm{i}}{2}\frac{d^{2}}{d\xi^{2}} -  \mathrm{i}E + \frac{\mathrm{i}}{2}\xi^{2} \right) \phi_{2} = 0 
 \end{equation}
which upon simplification yields 
 \begin{equation}
\left(-\frac{1}{2}\frac{d^{2}}{d\xi^{2}} - \frac{1}{2} \xi^{2}\right)\phi_{2} = -\hat E \phi_{2} 
\end{equation}
which is Eq.~(\ref{IHO_equation}) again.
\end{enumerate}
Thus, we have shown that both $\phi_{1}(\xi)$ and $\phi_{2}(\xi)$ are solutions of the IHO equation with same energy, and the claim is proved. Furthermore, these two solutions are linearly independent as their Wronskian is non-zero \cite{NIST:DLMF} 
\begin{equation}
\mathcal{W}\left(D_{\mathrm{i}\hat E - \frac{1}{2}}(\sqrt{2}e^{-\mathrm{i}3\pi/4}\xi),~D_{-\mathrm{i} \hat E -\frac{1}{2}}(\sqrt{2}e^{-\mathrm{i}\pi/4}\xi) \right) = -\mathrm{i}e^{-\pi \hat E/2} \, ,
\end{equation}
and so the solution given Eq.\ (\ref{pcfs1}) is indeed the general solution to the IHO Schr\"{o}dinger equation (\ref{ihoxipi}) with energy $-\hat{E}$.

 \section{The quantum canonical transform from wavefunctions in $Q$ to wavefunctions in $\xi$}
 \label{A1}
 
 In this appendix we explain the basic idea behind quantum canonical transforms and also derive the two specific transforms given in the main text in Eqns.\ (\ref{pcf_from_BK2}) and (\ref{second_pcf_from_BK}). The quantum canonical transforms used in this paper take wavefunctions in the $Q$ variable and map them to wavefunctions in the $\xi$ variable. Both the ISP and BK wavefunctions are functions of $Q$, whereas the IHO wavefunctions are functions of $\xi$.  It is perhaps surprising that two different transforms are needed, but this can ultimately be traced back to the fact that the BK wavefunctions live in two disconnected half-spaces as summarized in Eq.\ (\ref{Qrepwavefunction}). This means that  the
BK system is governed by a first order differential equation with apparently only a single solution whereas the ISP and IHO systems are governed by second order differential equations  with two independent solutions. To map between these different systems therefore requires some ingenuity; in reference \cite{brout1995primer} they solve the problem of mapping between the BK and IHO systems by obtaining their second solution from the momentum space representation of the BK Hamiltonian, whereas we prefer to remain in a single representation and instead solve the problem by squaring the BK Hamiltonian which has the added bonus of straightforwardly connecting to the ISP system. One interpretation of the squaring is that by allowing both energies $\pm \hat{E}$ it in some sense includes both particle and antiparticle type solutions  on an equal footing. Equivalently, we note that the BK Hamiltonian breaks time reversal symmetry since it is not invariant under $P \rightarrow -P$, but squaring restores this symmetry.

 \subsection{From Berry-Keating states to inverted harmonic oscillator states}
\label{QCT_BK_first_state}
References \cite{brout1995primer, ullinger2022logarithmic} describe the mapping of the BK states to parabolic cylinder states which we now discuss in detail.  The classical version of this mapping is a canonical transformation and it is worthwhile recalling the theory of canonical transformations via generating functions in classical mechanics \cite{goldstein2001}. There are four classes of generating functions but for the first transform [as given in Eq.\ (\ref{pcf_from_BK2})] we need the first class $F_{1}=F_{1}(\xi,Q,t)$. This generates a transformation via the relations $P = - \partial F_{1}/\partial Q$ and $\pi = \partial F_{1}/\partial \xi$ where ($\xi$,$\pi$) and ($Q$,$P$) are the old and new phase space variables, respectively. Taking 
\begin{equation}
F_{1}(\xi, Q) = -\frac{\xi^{2}}{2} + \sqrt{2}\xi Q - \frac{Q^{2}}{2} \ ,
\end{equation} 
it can be readily verified that $F_{1}$ gives the required BK $\leftrightarrow$ IHO canonical transformation $Q = \frac{\pi + \xi}{\sqrt{2}}, P = \frac{\pi - \xi}{\sqrt{2}}$.

Coming now to the quantum case, the BK Schr\"{o}dinger equation in the $Q$ representation is [Eq.\ (\ref{euler_operator_Qrep}) in the main text]
 \begin{equation}
 Q\frac{\partial \phi}{\partial Q}  = \left(-\mathrm{i}\hat E -\frac{1}{2}\right)\phi(Q) \  ,
 \end{equation}
 and its solution is given by
 \begin{equation}
 \phi(Q) = Q^{-\mathrm{i} \hat E -1/2 }  \ .  \label{BK_state_Q}
 \end{equation}
To map this to the $\xi$ representation we apply a \textit{quantum canonical transform} \cite{dirac1945analogy,leacock1983hamilton,kim1999canonical} 
\begin{equation}
\phi(\xi) = \int dQ~\inpro{\xi}{Q}\, \inpro{Q}{\phi} = \int_{0}^{\infty} dQ~ \underbrace{ e^{\mathrm{i}F_{1}(\xi, Q)}}_{\inpro{\xi}{Q}}\, \underbrace{\phi(Q)}_{\inpro{Q}{\phi}}  \ .
\end{equation}
The function $F_{1}(\xi, Q)$ in the kernel $\inpro{\xi}{Q} = e^{\rm i F_{1}(\xi, Q)/\hbar} $ can be derived following Dirac  \cite{dirac1945analogy} 
\begin{equation}
\left< \xi \right|\, \pi \,\left| Q \right> = -\rm i\hbar \frac{\partial }{\partial \xi} \inpro{\xi}{Q} =  \left(\frac{\partial F_{1}}{\partial \xi} \right) \inpro{\xi}{Q} = \left< \xi \right| \frac{\partial F_{1}}{\partial \xi} \left| Q\right> \\
 \end{equation}
 which implies 
 \begin{equation}
 \pi = \frac{\partial F_{1}}{\partial \xi} \ .
 \end{equation}
 Similarly,
 \begin{equation}
\left< \xi \right|\, P \,\left| Q \right> = \rm i \frac{\partial }{\partial Q}  \inpro{\xi}{Q} = -\frac{\partial F_{1}}{\partial Q}  \inpro{\xi}{Q} = - \left< \xi \right|\, \frac{\partial F_{1}}{\partial Q} \,\left| Q \right>\\
 \end{equation}
 which implies 
 \begin{equation}
 P = -\frac{\partial F_{1}}{\partial Q} \ .
 \end{equation}
 Together these yield  
 \begin{equation}
 \label{eq:kernel1}
 \inpro{\xi}{Q} = e^{\mathrm{i}F_{1}(\xi, Q)} = e^{\mathrm{i}\left(-\frac{\xi^{2}}{2} + \sqrt{2}\xi Q - \frac{Q^{2}}{2} \right) } \ .
 \end{equation}

 The quantum canonical transform to the $\xi$ representation is therefore
\begin{equation}
\phi(\xi)= \int_{0}^{\infty} dQ~Q^{-\mathrm{i} \hat E - 1/2} e^{\mathrm{i}\left(-\frac{\xi^{2}}{2} + \sqrt{2}\xi Q - \frac{Q^{2}}{2} \right)  }   \label{App:pcf_from_BK}
\end{equation}
which is also an example of a Mellin transform \cite{melintransform_xp}. Making use of the representation of the parabolic cylinder function $D_{s} (\xi) $ given in Eq.\ (B6) in reference \cite{ullinger2022logarithmic}  
 \begin{equation}
 \label{eq:PCF_integral_representation}
 D_{s}\left(\frac{b}{\sqrt{2 a}} \right) = \frac{\left(2a\right)^{-s/2}}{\Gamma(-s)} e^{- \frac{b^{2}}{8a}} \int_{0}^{\infty} dQ~ Q^{-s-1} e^{-aQ^{2} - bQ} \ ,
 \end{equation}
and putting 
\begin{eqnarray} 
a &=& \mathrm{i}/2  \label{eq:F1a} \\  s &=& \mathrm{i}\hat{E} - 1/2  \label{eq:F1s} \\ b &=& -\sqrt{2}\mathrm{i}\xi \label{eq:F1b}
\end{eqnarray}
 one recognizes that the wavefunction  $\phi(\xi)$ in Eq.\ (\ref{App:pcf_from_BK}) can be written in terms of parabolic cylinder functions as 
\begin{equation}
\label{eq:QCTalpha}
\phi(\xi)  =  e^{-\pi \hat E/4} e^{-\mathrm{i}\pi/8}  \, \Gamma\left(\frac{1}{2} - \mathrm{i}\hat E\right) \, D_{\mathrm{i}\hat E -1/2} \left(\sqrt{2} e^{-\mathrm{i}3\pi/4} \xi  \right) 
 \end{equation}
 which is Eq.\ (\ref{first_pcf_from_BK_explicit}) for $\phi_{1}(\xi)$ in the main text (apart from the coefficient $\alpha$).
 
 How do we find the second solution $\phi_{2}(\xi)$? At first sight it seems that there is only a single solution $\phi(Q)$, as given in Eq.\ (\ref{BK_state_Q}), available on the BK side to transform over to the IHO side. One way to get a second solution is to change into the momentum representation for the BK Schr\"{o}dinger equation
 \begin{equation}
 P\frac{\partial \psi}{\partial P}  = \left(\mathrm{i}\hat E -\frac{1}{2}\right)\psi(P) \  , 
 \end{equation}
 which has the solution $\psi(P)=P^{+\mathrm{i}\hat{E}-1/2}=[\phi(P)]^{\ast}$. This has the same structure as the $Q$-space solution but is complex conjugated \cite{berry1999h}. Complex conjugation effectively changes the sign of $\hat{E}$ and can be interpreted as a time reversal operation, as is also evident by comparing the classical solutions for the position and momentum variables given in Eq.\ (\ref{time_logQ}). A quantum canonical transformation of $\psi(P)$ with the kernel \cite{brout1995primer}
 \begin{equation}
 \label{eq:F2QCT}
 F_{2}(\xi, P) = \frac{\xi^{2}}{2} + \sqrt{2} \xi P + \frac{P^{2}}{2}
 \end{equation}
 gives the second IHO solution. As the notation indicates, $F_{2}(\xi,P)$ is a member of the second class of generating functions \cite{goldstein2001}.   
 
 However, in this paper we prefer to remain in the $Q$-representation and will not follow this route. Instead we shall show in the next section that when the BK Hamiltonian is squared (which is a step in the full mapping from ISP to IHO) a second spatial BK solution $Q^{+\mathrm{i}\hat{E}-1/2}$ becomes allowed because both energies $\pm \hat{E}$ give the same eigenvalue $\hat{E}^{2}$. This provides the second solution without the need to invoke the momentum space solution.

\subsection{From inverse square potential states  to inverted harmonic oscillator states} 
\label{second_state_details}
The zero energy ISP Schr\"{o}dinger equation [Eq.\   (\ref{inverse_sq_eqn}) in the main text] is 
\begin{equation}
\left(-\frac{\partial^{2}}{\partial Q^{2}} - \frac{\hat E^{2} + 1/4}{Q^{2}}\right)\chi(Q) = 0 \ ,
\end{equation}
where $-\hat{E}$ is the dimensionless energy of the IHO and BK Hamiltonians that here determines the depth of the ISP (we recall that the zero-energy ISP system is directly related to square of the BK system). If $\hat E \neq 0$ we have an unbounded-from-below (``super-critical'') ISP system, and there is an ambiguity in the boundary condition at $Q = 0$.
The general solution to the zero energy ISP Schr\"{o}dinger equation takes the form 
\begin{equation}
\chi(Q)  = \alpha Q^{\frac{1}{2}-\mathrm{i}\hat E} + \beta Q^{\frac{1}{2}+\mathrm{i}\hat E}     \label{inverse_square_states}
\end{equation}
which is Eq.\ (\ref{wavefunction_inverse_square}) in the main text. The two terms in this wavefunction are linearly independent with Wronskian non-zero if $\hat E \neq 0$.

\par  One can perform a quantum canonical transform integral directly from the ISP states to the IHO states $\phi(\xi)$ because the ISP states are in the same variable $Q$ as the BK states [according to Eq.\ (\ref{BKtoInvSqPsi}) the relationship between the BK states $\phi(Q)$ and the ISP states $\chi(Q)$ is $ \phi(Q) = \chi(Q) / Q $] which are in turn related to the IHO states by the canonical transformation discussed in \ref{QCT_BK_first_state}. However, the kernel needed in the quantum canonical transform is different for the two ISP states. For the first ISP state, $\alpha Q^{\frac{1}{2}-\mathrm{i}\hat E}$, which maps to the BK state $\alpha Q^{-\frac{1}{2} - \mathrm{i}\hat E}$,  the kernel has already been derived in \ref{QCT_BK_first_state}, namely  $e^{\mathrm{i} F_{1}(\xi, Q)}$, as given in Eq.\ (\ref{eq:kernel1}). Hence, the first ISP solution can be mapped directly to the first IHO solution as
 \begin{equation}
 \phi_{1}(\xi) = \int_{0}^{\infty} dQ~\alpha Q^{-\mathrm{i}\hat E - 1/2} e^{\mathrm{i}\left(-\frac{\xi^{2}}{2} + \sqrt{2}\xi Q - \frac{Q^{2}}{2} \right)  } \label{App:pcf_from_BK2}
 \end{equation}
 which upon using the integral representation of the parabolic cylinder function $D_{s} (\xi) $ given in Eq.\ (\ref{eq:PCF_integral_representation})
with $a = \mathrm{i}/2$, $s = \mathrm{i} \hat{E} - 1/2$, $b = -\sqrt{2}\mathrm{i}\xi$ yields 
\begin{equation}
\phi_{1}(\xi) = \alpha \, e^{- \pi \hat{E}/4}  e^{-\mathrm{i}\pi/8} \, \Gamma\left( \frac{1}{2} - \mathrm{i}\hat E\right) \, D_{\mathrm{i} \hat E - 1/2}\left(\sqrt{2} e^{-\mathrm{i}3\pi/4}\xi \right) \ ,
\end{equation}
like in Eq.\ (\ref{eq:QCTalpha}) above and also in Eq.\ (\ref{first_pcf_from_BK_explicit}) in the main text. The second ISP solution $\beta Q^{\mathrm{i} \hat E + 1/2}$ gives the second IHO solution via a different quantum canonical transform (to be derived below)
 \begin{equation}
 \phi_{2}(\xi) =  \int_{0}^{\infty} dQ~\beta Q^{\mathrm{i}\hat E - 1/2} e^{\mathrm{i} G (\xi,Q)}  = \int_{0}^{\infty} dQ~\beta Q^{\mathrm{i}\hat E - 1/2} e^{\mathrm{i}\left(\frac{\xi^{2}}{2} + \sqrt{2}\xi Q + \frac{Q^{2}}{2} \right)} \ . \label{second_pcf_from_BK_appendix}
 \end{equation}
As before, we can use the integral representation of the parabolic cylinder function in Eq.~(\ref{eq:PCF_integral_representation}) to write this integral in terms of $D_{s}(\xi)$ as 
\begin{equation}
\phi_{2}(\xi) = \beta e^{- \pi \hat{E}/4} e^{\mathrm{i}\pi/8} \, \Gamma\left(\frac{1}{2} + \mathrm{i}\hat E\right) \,  D_{-\mathrm{i}\hat E - 1/2}(\sqrt{2}e^{-\mathrm{i}\pi/4}\xi)
\end{equation}
where this time we have put $a = -\mathrm{i}/2$, $s = - \mathrm{i} \hat E - 1/2$, $b = -\sqrt{2}\mathrm{i}\xi$. This is Eq.\ (\ref{second_pcf_from_BK_explicit}) in the main text.
 To derive the generating function $G(\xi,Q)=\xi^{2}/2 + \sqrt{2}\xi Q + Q^{2}/2$ used in the second quantum canonical transform we can use the following symmetry of the squared Berry-Keating equation: $Q \rightarrow P, ~ P\rightarrow -Q$. This transformation is canonical because it preserves the commutation relation $[Q, P] = \mathrm{i}$. Under this transformation the BK Hamiltonian transforms as  $H_{\mathrm{BK}} \rightarrow -H_{\mathrm{BK}}$, and we note the second ``BK state'' $\beta \,  Q^{\mathrm{i}\hat E-1/2}$ is an eigenfunction $-H_{\mathrm{BK}}$.  Because of the squaring step this transformation is a symmetry of the squared BK equation of motion.  Hence one can use the generating function $G(\xi, Q) = \frac{\xi^{2}}{2} + \sqrt{2}\xi Q + \frac{Q^{2}}{2}$ that generates the following canonical transformation 
\begin{eqnarray}
Q \rightarrow P \quad\mathrm{implies}&\quad Q = \frac{\pi - \xi}{\sqrt{2}}\\
P \rightarrow -Q \quad\mathrm{implies}&\quad P = -\left(\frac{\pi + \xi}{\sqrt{2}}\right)
\end{eqnarray} 
to build the second linearly independent IHO state using the squared BK state. Comparing the function $G(\xi,Q)$ with $F_{2}(\xi,P)$ in Eq.\ (\ref{eq:F2QCT}) we see they have an identical structure except for exchanging $Q$ and $P$.

 Finally we note that the Wronskian for these two parabolic cylinder states is non-zero \cite{NIST:DLMF} 
\begin{equation}
\mathcal{W}\left(D_{\mathrm{i}\hat E - 1/2}(\sqrt{2}e^{-\mathrm{i}3\pi/4}\xi),~D_{-\mathrm{i}\hat E -1/2}(\sqrt{2}e^{-\mathrm{i}\pi/4}\xi) \right) = -\mathrm{i}e^{-\pi \hat E/2}
\end{equation}
 which means they are linearly independent. So we get two linearly independent solutions (parabolic cylinder functions) to the IHO in the $\xi$ variables from the inverse square solutions in the $Q$ variables. 
 
 \section{Solutions to the RG differential equation for the IHO system}
\label{A2}

In the main body of the paper, Eq.~(\ref{betafunction_IHO}) gives the differential equation for the RG flow for the IHO system 
\begin{equation}
L\frac{\exd \Lambda_{\rm IHO}}{\exd L} =  \left( \frac{m\omega L^{2}}{\hbar} - \hat E \right) \left(\Lambda_{\rm IHO}^{2} + 1 \right) \ .
\end{equation}
By integrating the above equation for a given initial condition $\Lambda_{0}$ we find  the solution 
\begin{equation}
\Lambda_{\rm IHO} = \frac{ \Lambda_{0} + \tan\left(m\omega/2\hbar\, L^{2} - \hat E\ln(\sqrt{2m\omega/\hbar}L) \right) }{ 1  - \Lambda_{0}\tan\left(m\omega/2\hbar\, L^{2} - \hat E\ln(\sqrt{2m\omega/\hbar}L) \right) } \ .
\end{equation}
The plot given in \fref{fig6} shows the real and imaginary parts of this solution as a function of $\ln(\sqrt{m\omega/\hbar}L)$. The phase portrait in the complex plane is given in \fref{fig7} which exhibits limit cycles like in the ISP system, however the flow of the IHO system is faster than the log periodic behaviour found in the ISP and is discussed in detail in \Sref{sec:mapping}.  
 
\section*{References}
\bibliographystyle{iopart-num}

\providecommand{\newblock}{}

\end{document}